# MAPPING THE FUTURE OF HUMAN DIGITAL TWIN ADOPTION IN JOB-SHOP INDUSTRIES: A STRATEGIC PRIORITIZATION FRAMEWORK

**Samiran Sardar**[1], **Nasif Morshed**[2*] and **Shezan Ahmed**[3]

Department of Industrial Engineering and Management
Khulna University of Engineering and Technology, Khulna 9203, Bangladesh
samirankuet40@gmail.com[1], morshed1911041@gmail.com[2*] and shezan.kuetipe@gmail.com[3]

***Abstract*** - *Although Digital Twin is actively deployed in manufacturing, its human-centric counterpart - Human Digital Twin (HDT) is understudied, especially in job-shop production with high task variability and manual labor. HDT applications like ergonomic posture monitoring, fatigue prediction and health-based task assignment offer benefits to industries in emerging economies. However, poor digital maturity, lack of awareness and doubts about use-case applicability hinder adoption. This study provides a strategic prioritization framework to aid human-centric digital evolution in labor-intensive industries for guiding the selection of HDT applications delivering the highest value with the lowest implementation threshold. An integrated Fuzzy AHP-TOPSIS approach evaluates the use-cases based on criteria like implementation cost, technological maturity, scalability. These criteria and use-cases were identified based on input from a five-member expert panel and verified for consistency (CR < 0.1). Analysis shows posture monitoring and fatigue prediction as most influential and practicable, especially in semi-digital environments. Strengths include compliance with Industry 5.0 principles incorporating technology and human factors. Lack of field validation and subjective knowledge pose drawbacks. Future work should include simulation-based validation and pilot tests on real job-shop settings. Ultimately, the research offers a decision-support system helping industries balance innovativeness and practicability in early stage of HDT adoption.*

**Keywords:** Human Digital Twin, Job-Shop Industries, Fuzzy AHP-TOPSIS, Use-Case Prioritization, Industry 5.0.

## 1. INTRODUCTION

The rapid development of digital technologies has positioned the Human Digital Twin (HDT) as a critical focus within industrial manufacturing, particularly in job-shop. HDT provides real-time insights, digitally mirroring workers' cognitive and physical states to optimize performance. HDTs offer the capability to increase worker safety, streamline processes and dynamically adapt accordingly to change in job-shop managements, which are known to be custom and flexible in character. Adoption faces barriers including high cost, data integration challenges and complexity of simulating human behavior [1]. Successful HDT implementation requires operations management to clearly define strategic objectives, rigorously evaluate existing data infrastructure and align HDT technology selection with current operational requirements and long-term organizational goals [1], [2].

The effective HDT project starts with a well-defined phase to identify high-value workstations and processes where digital mirroring can provide the most significant operational benefit. Social factors (along with technical and operating factors) testify to the need of a thorough framework through which to consider and prioritize paths towards the implementing of HDT [2]. These include several advantages such as predictive maintenance, enhanced productivity and effective control of the workforce - all contributing to the success of creating a resilient and future-ready operation, which are optimized through a tactical path towards adoption of HDT by job-shop [3]. However, the process of successful HDT implementation on the road is not a common solution. Job shops must be competitive with production requirements, raw material availability and manpower issues that are unique to each shop [3]. This implies that the choice of HDT use cases and prioritization should be based on an explicit knowledge of organization priorities as well as technological receptiveness and organizational involvement. Indicators such as productivity, downtime reduction and employee satisfaction facilitate ongoing HDT improvement and demonstrate its real value in digital transformation.

Against this backdrop, the primary objectives of this study are operationalized through the following research questions:

**RQ1.** What are the main challenges and enablers that affect the adoption of HDT in job-shop industries?

**RQ2.** How can manufacturers choose and prioritize HDT use-cases that are most beneficial in terms of safety, productivity and adaptability?

Through addressing such questions, this study will not only offer a practical guideline to manufacturers that are keen to tackle their HDT journey but a strong, research-based framework of prioritizing and actualizing the potential of HDT within job-shop environment by employing a hybrid Multi-Criteria Decision Making (MCDM) method - Fuzzy AHP and TOPSIS.

## 2. LITERATURE REVIEW

The evolution of digital technologies has driven the adoption of digital twins (DTs) and, more recently, human digital twins (HDTs) in manufacturing. HDTs are increasingly recognized for their ability to enhance safety, optimize performance and inform decision-making through real-time data analysis [4], [5]. In aerospace manufacturing, HDTs can improve quality assurance by enabling interactive monitoring between operators and machines [4]. Safety-focused HDT applications such as posture detection, real-time fatigue prediction and PPE tracking can significantly reduce workplace accidents and improve compliance, thereby enhancing safety analysis and risk management in job-shop environments [6], [7]. Beyond safety, HDTs can also enable health-based task assignment, where worker health data guides workload distribution [8], and skill training simulation, where immersive digital environments support workforce development and adaptive training [9]. Despite their potential, HDTs face significant technological challenges. Many job-shop implementations remain at an early maturity stage, advancing only slowly from descriptive monitoring toward predictive and prescriptive capabilities [10]. Scalable solutions require interoperable methods of gathering shop-floor data, yet fragmentation of existing systems complicates this [11]. Managing heterogeneous data from sensors, machines and human activities further increases complexity, requiring robust approaches such as graph-based uncertainty management [12]. Organizational barriers also hinder adoption. High initial costs related to hardware, software and workforce training remain the greatest impediment, especially for SMEs with limited resources [13]. Furthermore, high-fidelity HDTs demand sophisticated data models and secure, scalable platforms, adding to financial and technical burdens [12]. Collectively, these themes emphasize that the strategic adoption of HDTs in job-shop industries relies critically on assessing technological maturity, implementation cost, scalability and the organizational capacity to manage complex, heterogeneous data landscapes. While these studies provide valuable insights into individual enablers and barriers, there remains a lack of integrated prioritization methods suited to heterogeneous job-shop settings.

The selection of the Fuzzy AHP-TOPSIS integrated framework is justified by its inherent ability to effectively handle the qualitative and uncertain nature of expert decision-making [14], [15]. Fuzzy AHP is superior for weighting criteria as it accurately transforms linguistic expert judgments into quantifiable fuzzy numbers, which is crucial when dealing with subjective criteria like 'Safety Impact' and 'Technological Maturity' [15]. The TOPSIS method is then employed for ranking because it provides a clear, compensatory mechanism by evaluating the distance of each alternative from both the ideal best and ideal worst solutions [14]. While alternative MCDM methods like Fuzzy VIKOR are available, the combined strength of FAHP for robust criteria weighting and TOPSIS for definitive ranking makes it highly appropriate for strategic prioritization in resource-constrained environments like job-shops [16].

## 3. METHODOLOGY
### 3.1 Problem Definition and General Methodology

The adoption of Human Digital Twin (HDT) technologies into job-shop conditions has become one of the key enablers of the industry 4.0 shift as it enables the improvement of the flexibility of operations, safety of workers, proactive maintenance of equipment, and the overall process of decision-making. However, the strategy of implementation suffers because of a list of substantive barriers. First, the lack of standardization of prioritization framework limits the managerial ability to identify and pick the most relevant HDT use-cases. Second, managerial discretion is often relevant to decision making thereby contributing to resources misallocation and unevenly distributed intervention. Finally, the evaluation of HDT uptake should consider concurrently the quantitative measure, which includes cost, efficiency, and sustainability as well as the qualitative measure that includes the organizational preparedness, and workforce acceptance, hence throwing a lot of uncertainty to the decision-making [14], [15].

This study proposes one of the hybrids multi-criteria decision making (MCDM) models that merges Fuzzy Analytic Hierarchy Process (FAHP) and Technique of Order of Preference by Similarity to Ideal Solution (TOPSIS). FAHP provides a powerful tool in making valid criterion weightings under non-certainty or verbal expression of the expert judgment. This is further boosted by TOPSIS technique that allows ranking of HDT adoption use-cases among optimal and worst optimal solutions, and hence captures both qualitative and quantifiable factors. The methodological framework is shown in Fig. 1.

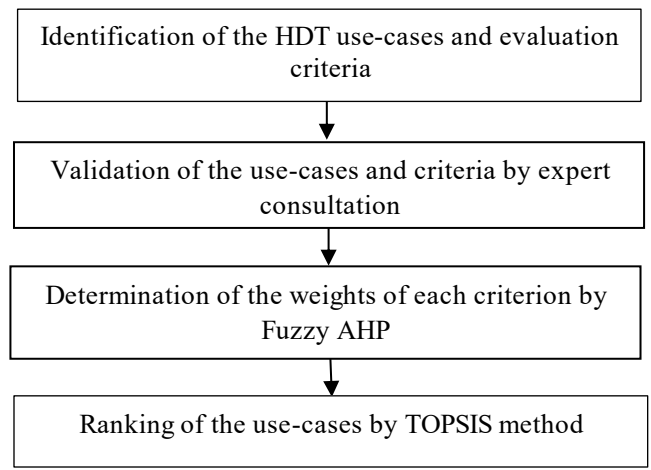

Fig. 1 The methodological framework

Through the first stage, a systematic review of the literature was conducted to find out the possible use-cases of hardware-driven technology (HDT) and the performance metrics relevant to job-shop industries. The review found out five main evaluation criteria: Safety Impact, Technological Maturity, Implementation Cost, Data Requirement Complexity, and Scalability. The criteria were then cross-linked to the practical HDT applications which include Posture Monitoring, Fatigue Prediction, PPE Compliance Tracking, Health-Based Task Assignment and Skill Training Simulation. After the literature review, expert survey was carried out to verify and validate the selection. This stage entailed the participation of a panel made up of five experts that included academic research and industry practitioners. The panel was specifically chosen to ensure a balanced perspective, with three members having deep theoretical knowledge of HDT/MCDM and two having over ten years of practical experience in job-shop operations management. Individual experts were requested to assess the applicability of proposed use-cases and criteria based on a five point-Likert scale. The finalized list was confirmed based on their response's concurrence and below-average confidence and eliminated items that had a low consensus or unimportant as shown in Table 1. This has provided an assurance that the assessment has incorporated both theoretical knowledge based on academic research and practical views based on industry.

At the second phase, FAHP was used to estimate the relative weight (importance) of the chosen evaluation criteria. Pairwise comparisons of the criteria by linguistic judgements in primary form were given by experts and expressed in the form of Triangular Fuzzy Numbers (TFNs) to effectively capture and manage the inherent vagueness and imprecision in decision-making.

In the third phase, the TOPSIS algorithm was used to rank HDT use-cases. Criteria values were compared to each HDT alternative using the weights produced in FAHP and the proximity coefficients were made in ranking the alternatives to the positive ideal solution (PIS) and the negative ideal solution (NIS). Utilizing the combined FAHP-TOPSIS framework, the study brings out the twin problems of uncertainty of the expert judgment and lack of an organized prioritization process of HDT adoption. It, therefore, provides a solid guide to managers and policy makers who work in industries that are governed by job shops.

Table 1. HDT Evaluation Criteria and Use-Cases

| Evaluation Criteria | Use-cases |
|---|---|
| Safety Impact | Posture Monitoring |
| Technological Maturity | Skill Training Simulation |
| Implementation Cost | Fatigue Prediction |
| Data Requirement Complexity | Health-Based Task Assignment |
| Scalability | PPE Compliance Tracking |

### 3.2 Methodology for the Application of AHP Technique

Analytic Hierarchy Process (AHP) is a multi-criteria decision-making technique that is aimed at solving a problem by breaking it down into its component solutions, cluster these solutions and place them on a hierarchical scale [16]. A fuzzy triangular scale is also determined using fuzzy logic to convert the scale of AHP to a triangular scale that is fuzzy hence simplifying the establishment of the values of priorities. The fuzzy AHP process can be outlined as the following steps:

**Step 1.** First, the problem purpose and its hierarchical structure was developed.
**Step 2.** To measure relative significance of an individual indicator of each enabler, linguistic variables were constructed. The indicators are shown in Table 2.
**Step 3.** The pairwise comparisons of criteria were provided by experts in the form of linguistic variables (e.g., as important, more important moderately, very strongly important). These evaluations are in turn converted to Triangular Fuzzy Numbers (TFNs) that have the form:
$$A = (l, m, u)$$
where $l, m,$ and $u$ denote the lower, middle and upper bounds of the expert judgement.
**Step 4.** The fuzzy pairwise matrix was constructed for each category of the enablers based on the expert's review.
**Step 5.** After creating a pairwise comparison matrix, the consistency of each respective fuzzy matrix was measured to ensure the validity of the judgment that was made by the experts. The geometric mean method and α-cut approach were used to check consistency and the final aggregated matrix demonstrated a Consistency Ratio (CR) of 0.08, which is below the acceptable threshold of 0.1, confirming the reliability of the expert judgments. If the TFN is represented by $a_{ij} = (l_{ij}, m_{ij}, u_{ij})$, then the fuzzy geometric mean is calculated by Eq. (1). The fuzzy geometric mean of row $i$. for each criterion i, the fuzzy geometric mean value was calculated as:

$$r_i = \left( \left( \prod l_{ij} \right)^{\frac{1}{n}}, \left( \prod m_{ij} \right)^{\frac{1}{n}}, \left( \prod n_{ij} \right)^{\frac{1}{n}} \right) \quad (1)$$

where,
$r_i$ = fuzzy geometric mean of row i,
$l_{ij}, m_{ij}, u_{ij}$ = lower, middle and upper values of TFN,
n = number of criteria

**Step 6.** After completing fuzzy geometric mean, then the fuzzy weights and normalized weights were calculated by Eq. (2) and Eq. (3). The fuzzy weights were obtained by normalizing the fuzzy geometric means:

$$w_i = r_i \otimes (r_1 \oplus r_2 \oplus \dots, r_n)^{-1} \quad (2)$$

In expanded form:
$$w_i = \left( \frac{l_i}{\sum u_k}, \frac{m_i}{\sum m_k}, \frac{u_i}{\sum l_k} \right)$$

where,
$\otimes$ = fuzzy multiplication
$\oplus$ = fuzzy addition
$(l, m, u)^{-1} = (1/u, 1/m, 1/l)$

To convert fuzzy weights into crips values for ranking, the center of gravity method was used:

$$w_i^* = \frac{(l_i' + m_i' + u_i')}{3} \quad (3)$$

where,
$w_i^*$ = normalized weights
$(l_i', m_i', u_i')$ are the fuzzy weight components of criterion i.

Table 2. Linguistic variables and equivalent Numerical Crisp values and TFNs.

| Linguistic Variable | TFNs |
|---|---|
| Equally important | (1, 1, 1) |
| Moderate important | (2, 3, 4) |
| Strong important | (4, 5, 6) |
| Very strong important | (6, 7, 8) |
| Extremely important | (9, 9, 9) |
| Equally important to moderately more important | (1, 2, 3) |
| Moderately more important to strongly more important | (3, 4, 5) |
| Strongly to very strongly more important | (5, 6, 7) |
| Very strongly to extremely more important | (7, 8, 9) |

## 3.3 Methodology for the Application of TOPSIS Technique

TOPSIS is used to address multi-criteria decision making (MCDM) issues and to select the most appropriate alternative, which involves the calculation of the shortest distance to the positive ideal solution ($A^+$) and the longest distance from the negative ideal solution ($A^-$). The closeness to the positive ideal solution determines the maximum functionality and distance to the negative ideal solution determines the minimum functionality. TOPSIS steps are:

**Step 1.** Development of a decision matrix based on professional judgement. Let $A = \{A_1, A_2, \ldots, A_m\}$ represent HDT use-cases and $C = \{C_1, C_2, \ldots, C_n\}$ the evaluation criteria. The decision matrix is given in Eq. (4).

$$D = [x_{ij}]_{m \times n} \quad (4)$$

where, $x_{ij}$ is the performance of alternative $A_i$ with respect to criterion $C_j$.

**Step 2.** The decision matrix was normalized by Eq. (5).

$$r_{ij} = \frac{x_{ij}}{\sqrt{\sum_{i=1}^{m} x_{ij}^2}}, \text{ where } i = 1, 2, \ldots, m \text{ and } j = 1, 2, \ldots, n$$

**Step 3.** The weighted normalized matrix was derived by multiplying the normalized decision matrix and the weight vector respectively. This correlation is indicated in Eq. (6).

$$v_{ij} = w_j \times r_{ij} \quad (6)$$

**Step 4.** The positive ideal solution (PIS) and negative ideal solution (NIS) were determined by Eq. (7) and Eq. (8)

For a positive ideal solution

$$A^+ = \{max\ v_{ij} \mid j \in J_b\ ;\ min\ v_{ij} \mid j \in j_c\} \quad (7)$$

For a negative ideal solution

$$A^- = \{min\ v_{ij} \mid j \in J_b\ ;\ max\ v_{ij} \mid j \in j_c\} \quad (8)$$

where $J_b$ and $j_c$ represent benefit and cost criteria sets, respectively.

**Step 5.** Calculation of Euclidean distance of each alternative from positive ideal solution and negative ideal solution was computed by Eq. (9) and Eq. (10).

$$d_i^+ = \sqrt{\sum_{j=1}^{n} (v_{ij} - A_j^+)^2} \quad (9)$$

$$d_i^- = \sqrt{\sum_{j=1}^{n} (v_{ij} - A_j^-)^2} \quad (10)$$

**Step 6.** After completing the Euclidean distance, the relative closeness of each alternative was calculated by Eq. (11)

$$CC_i = \frac{d_i^-}{d_i^+ + d_i^-}, \quad 0 \leq CC_i \leq 1 \quad (11)$$

where $i = 1, 2, \ldots, m$.

**Step 7.** Based on the value of $C_i$, the ranking of alternatives was done to select the best one.

## 4. RESULTS AND DISCUSSION

The relative importance of evaluation criteria was obtained by using Fuzzy Analytic Hierarchy Process (FAHP) analysis. Table 3 indicates that the two most important factors of the five are Technological Maturity (0.352) and Safety Impact (0.343), whose weights add up to nearly 70%. Implementation Cost (0.152) is therefore placed as a second-order factor, with Data Requirement Complexity (0.095) and Scalability (0.057) having been deemed to be lower impact. These results suggest that job-shop industries consider adoption pathways that include safeguards for reliable technology readiness and the safety of workers first and foremost, with cost and scalability concerns following. The prominence of safety and maturity is agreeable with previous Industry 4.0 adoption studies where enablers including AI readiness, ERP, and compliance technologies were consistently ranked above cost-driven considerations [15], [16].

Table 3. Normalized Fuzzy AHP Weights of Criteria

| Criteria | Normalized Weight | Rank |
|---|---|---|
| Safety Impact | 0.343 | 2 |
| Technological Maturity | 0.352 | 1 |
| Implementation Cost | 0.152 | 3 |
| Data Requirement Complexity | 0.095 | 4 |
| Scalability | 0.057 | 5 |

Fig. 2 shows the distribution of weights, showing the dominance of Technological Maturity and Safety Impact. This distribution is consistent with the premise that safety-enhancing and technologically proven solutions are the major drivers of early HDT adoption.

Using the obtained weights, the Technique for Order of Preference by Similarity to Ideal Solution (TOPSIS) was applied to rank the identified use-cases. The decision matrix which was formed by expert evaluations is given in Table 4 while the weighted normalized matrix is given in Table 5.

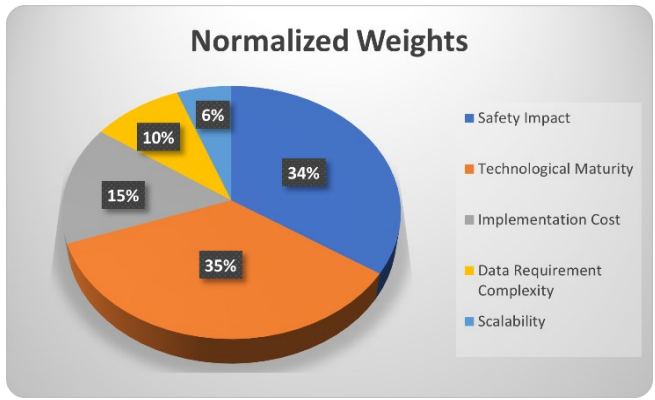

Fig. 2 Normalized weights of each criterion.

The Positive Ideal Solution ($A^+$) and Negative Ideal Solution ($A^-$) were found respectively as shown in Table 6. Over the entire period, the highest performance was consistently achieved by the Posture Monitoring dimension in both safety and maturity, while the Skill Training Simulation dimension was the one with the lowest contribution.

Table 4. Decision Matrix of HDT use-cases

| Use-Case \ Criteria | Safety Impact | Technological Maturity | Implementation Cost | Data Requirement Complexity | Scalability |
|---|---|---|---|---|---|
| Posture Monitoring | 8 | 8 | 7 | 5 | 8 |
| Skill Training Simulation | 7 | 7 | 6 | 6 | 7 |
| Fatigue Prediction | 9 | 7 | 5 | 7 | 7 |
| Health-Based Task Assignment | 8 | 6 | 6 | 8 | 6 |
| PPE Compliance Tracking | 6 | 9 | 8 | 8 | 9 |

Table 5. Weighted Normalized Decision Matrix

| Use-Case \ Criteria | Safety Impact | Technological Maturity | Implementation Cost | Data Requirement Complexity | Scalability |
|---|---|---|---|---|---|
| Posture Monitoring | 0.16 | 0.17 | 0.07 | 0.03 | 0.03 |
| Skill Training Simulation | 0.14 | 0.15 | 0.06 | 0.04 | 0.02 |
| Fatigue Prediction | 0.18 | 0.15 | 0.05 | 0.04 | 0.02 |
| Health-Based Task Assignment | 0.16 | 0.13 | 0.06 | 0.05 | 0.02 |
| PPE Compliance Tracking | 0.12 | 0.19 | 0.08 | 0.05 | 0.03 |

While the FAHP method used was verified for consistency, it is important to acknowledge that the criteria weights are derived from subjective expert judgment, representing a potential limitation. Future work should consider conducting a sensitivity analysis on the weight distribution to validate the robustness of the final ranking.

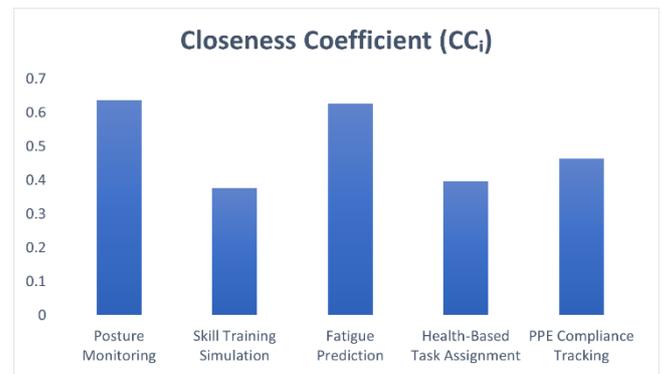

Fig. 3 Closeness Coefficients of HDT Use-Cases.

Table 6. Ideal Solutions

| Criteria | Type | Positive Ideal ($A^+$) | Negative Ideal ($A^-$) |
|---|---|---|---|
| Safety Impact | Benefit | 0.18 | 0.12 |
| Technological Maturity | Benefit | 0.19 | 0.13 |
| Implementation Cost | Cost | 0.05 | 0.08 |
| Data Requirement Complexity | Cost | 0.03 | 0.05 |
| Scalability | Benefit | 0.03 | 0.02 |

The distances to $A^+$ and $A^-$ and the closeness coefficients (CC) were then calculated, as shown in Table 7.

Table 7. TOPSIS Distances and Closeness Coefficients

| Use-Case | $d^+$ | $d^-$ | CC | Rank |
|---|---|---|---|---|
| Posture Monitoring | 0.035 | 0.062 | 0.639 | 1 |
| Skill Training Simulation | 0.059 | 0.036 | 0.379 | 5 |
| Fatigue Prediction | 0.042 | 0.071 | 0.628 | 2 |
| Health-Based Task Assignment | 0.068 | 0.045 | 0.398 | 4 |
| PPE Compliance Tracking | 0.07 | 0.061 | 0.466 | 3 |

As an illustration, the closeness coefficient for Posture Monitoring was computed as:

$$d^+ = 0.035, d^- = 0.062, CC = \frac{0.062}{0.035 + 0.062} = 0.639$$

This number validates Posture Monitoring as the highest-priority use-case. All the sorted options are also provided graphically in Fig. 3, where a clear distinction can be made between high-priority and low-priority applications.

The results show that safety-oriented and technologically mature HDT applications are most strategically relevant for job-shop sectors. Posture Monitoring, Fatigue Prediction and PPE Compliance Tracking come out on top of the list as they directly relate to improving occupational safety, ergonomics, and compliance (where the benefits are more immediate). On the other hand, the other two interventions, Health-Based Task Assignment and Skill Training Simulation, while important in the long-term, were lower priorities because of their relatively low immediate operational effect. These results not only confirm the primacy of safety and maturity, but also justify a staged adoption roadmap (Fig. 4). In the short-term, job-shops should make investments in posture monitoring; in the medium-term, investments in fatigue prediction and PPE compliance; and in the long-term, investments should be made in wider workforce development through Health-based task assignment and skill training simulation. This stepwise strategy will provide early wins whilst steadily developing long-term resilience.

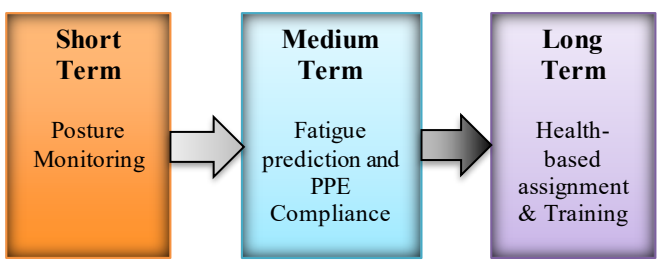

Fig. 4 Strategic adoption roadmap for HDT use-cases in job-shop industries.

The findings align with the previous FAHP-TOPSIS research like inventory categorization and Industry 4.0 adoption [16]. We also found that safety, readiness and maturity prevail over cost considerations in technology prioritization. These insights offer practical advice to managers: start with safe, safety-improving HDTs to build early trust and buy-in and then evolve toward more complex, capability-building uses. Collectively, the results imply that the adoption of HDT in job-shop industries should proceed in a safety-first and maturity-based manner, progressing toward an increasingly sophisticated human-digital integration over time as organizational readiness develops.

## 5. CONCLUSION

This study developed a strategic prioritization framework for Human Digital Twin (HDT) adoption in job-shop industries using a hybrid Fuzzy AHP-TOPSIS approach, revealing that posture monitoring and fatigue prediction emerge as the most influential and practical use-cases, while PPE compliance tracking, health-based task assignment and skill training simulation hold longer-term potential. The findings highlight the significance of prioritizing safety and technological maturity over cost and scalability, offering managers a stepwise roadmap for gradual yet impactful HDT integration. By aligning with Industry 5.0 principles, the framework not only advances worker-centric digitalization but also provides industries in emerging economies with actionable guidance on balancing innovation and practicality. However, the reliance on expert judgments and the lack of field validation remain limitations, underscoring the need for simulation-based verification and pilot testing in real shop-floor conditions. Future research should expand on larger-scale empirical studies, explore interoperability challenges, and test the framework across diverse industrial contexts to strengthen its generalizability and effectiveness. Ultimately, this research offers a structured pathway for job-shop industries to adopt Human Digital Twins in a safety-first and maturity-driven manner, paving the way for resilient, worker-centric digital transformation.